\documentclass[preprint,12pt]{elsarticle}

\usepackage{graphicx}
\usepackage{amssymb}
 \usepackage{amsthm}

\usepackage{lineno}
\usepackage{amsmath}
\newtheorem{theorem}{Theorem}
\newtheorem{lemma}{Lemma}

\begin{document}
	\begin{center}
	\large \bf	Empirical likelihood inference for partial functional linear regression models based on B-spline
	\end{center}
	
	\vspace*{1\baselineskip} 
	
	\begin{center}
		 Mingao Yuan and Yue Zhang\\
			\vspace*{1\baselineskip} 
		\textit{Department of mathematical science,}\\
		\textit{Indiana University-Purdue University Indianapolis,}\\
		\textit{Indianapolis, IN 46202, USA}\\
		E-mail:  mingyuan@iupui.edu,\ \ yz65@umail.iu.edu
		
	\end{center}

\begin{center}	
\today	
\end{center}

	\vspace*{1\baselineskip} 
	
\begin{center}	
	\bf Abstract	
\end{center}

In this paper, we apply empirical likelihood method to inference for the regression parameters in the partial functional linear regression models based on B-spline. We prove that the empirical log-likelihood ratio for the regression parameters converges in law to a weighted sum of independent chi-square distributions. Our simulation shows that the proposed empirical likelihood method produces more accurate confidence regions in terms of coverage probability than the asymptotic normality method.


	\section{Introduction}
	
	With the rapid development of measurement apparatus and computers, it is possible that the data are collected over an entire time period. In the literature, this kind of data are called \textit{functional data}. One of the goal of functional data analysis is to explain the variations of a dependent variable by using an independent functional variable. Functional linear model is one of the most popular models to realize this. There are mainly two approaches for estimation and inference in functional linear analysis. One of them is the functional principle component analysis(FPCA), which has the advantages of interpretability and availability of a good estimate of the slope function(Cai and Hall(2006), Hall and Horowit (2007), Shin (2009), Yuan and Cai (2010), Cai and Yuan (2012)). Another approach is the polynomial spline method. As a commonly used method in nonparametric or semiparametric analysis, the polynomial spline method has been introduced into functional data analysis by different authors(Ramsay and Silverman(1997), Cardot et al. (2003, 2005)). Ramsay and Silverman (1997,2005) and Cardot et al.(2003) employed spline to estimate the functional slope and Cardot et al.(2005) proposed a spline estimator for the functional coefficient in quantile regression.

	 To improve the accuracy of prediction and make the functional linear model more interpretable, sometimes other predictor variables should be incorporated into the model. Especially, Zhang et al. (2007) introduced the \textit{partial functional linear model} and applied it to analyze the effect of women's hormone on the total hip bone mineral density. Later, using functional principle component analysis, Shin (2009) proposed a new estimator for the parameters. For estimators based on B-spline in partial functional linear model, Zhou et al. (2016) established the asymptotic normality for the regression parameters and the global convergence rate for the slope function.

	 The empirical likelihood(EL) was introduced by Owen(1990, 2001) to construct confidence region in a nonparametric setting. As an analog of the parametric likelihood method, it has been extensively applied to different fields due to some of the nice properties. As nonparametric method, it doesn't require a prespecified distribution for the data. The confidence region respects the range of the data and usually performs better than that based on asymptotic normality. Recently, the EL method has been used for inferences under different models(Zhao (2010), Cheng et al.(2012)). 
	 In this paper, we propose the empirical likelihood based confidence region for the regression parameters in partial functional linear model and compare it with the ones based on asymptotic normality proposed in Zhou et al. (2016).

	The paper is organized as follows. Section 2 presents the partial functional linear model and the asymptotic normality based confidence region. In section 3, we develop the empirical likelihood confidence regions for the regression parameters. Section 4 includes simulation studies to illustrate the advantage of the EL based confidence region over asymptotic normality based confidence region in terms of coverage probability. The proof is presented in Section 5.

	\section{Asymptotic normality method}
	
	The partial functional linear model is 
	
	\[Y={\bf Z}^T{\bf\beta}+\int_0^1X(t)\alpha(t)dt+\epsilon,\]
	where $Y$ is scalar response variable, ${\bf Z}$ is $p$-dimensional predictor variable and $X(t)$ is a random process in $L^2([0,1])$. We assume that the random error $\epsilon$ is independent of $X(t)$ and ${\bf Z}$, $E(\epsilon)=0$ and $Var(\epsilon)=\sigma^2$. The $\beta$ is unknown $p$-dimensional parameter vector and $\alpha(t)$ is an unknown slope function in $L^2([0,1])$.	
	 Denote $||\phi_1||=(\int_0^1\phi_1^2dt)^{\frac{1}{2}}$ and $<\phi_1, \phi_2>=\int_0^1\phi_1(t)\phi_2(t)dt$, for $\phi_1,\phi_2\in L^2([0,1])$.
	 
		Let $0=t_0<t_1<\dots<t_{N_n}<t_{N_n+1}=1$ be a knot sequence in $[0,1]$. A polynomial spline of degree $k(k\geq 0)$ is a function such that it is a polynomial of degree $k$ on each interval $[t_j,t_{j+1}](j=0,1,\dots,N_n)$ and is $k-1$ times continuously differentiable in $[0,1]$. Let $S_{k,N_n}$ be the linear space spanned by splines with degree $k$ and number of knots $N_n$. It is well known that $S_{k,N_n}$ is of dimension $k_n=N_n+k+1$. Let $B_1,\dots, B_{k_n}$ be the B-spline basis. For $\alpha(t)\in C^{k+1}([0,1])$, we can approximate it by a unique linear combination of splines, that is,
	\[\alpha(t)\approx\sum_{i=1}^{k_n}b_kB_k.\]
	
	Let $(X_i, {\bf Z}_i, Y_i), i=1,\dots,n$ be the data from the model. Then the model can be written approximately as 
	
	\[Y_i={\bf Z}_i^T\beta+\sum_{s=1}^{k_n}b_s<X_s,B_s>+\epsilon_i,\ \ i=1,\dots,n.\]	
	Then the least square estimators of ${\bf b}=(b_1,\dots,b_{k_n})^T$ and $\beta$ are the minimizers of the following loss function
	
	\[\sum_{i=1}^n\big (Y_i-{\bf Z}_i^T\beta-\sum_{s=1}^{k_n}b_s<X_s,B_s>\big )^2.\]
	
	The profile least square estimators are given by	
	\[\hat{\beta}=({\bf Z}^T(I-A){\bf Z})^{-1}{\bf Z}^T(I-A){\bf Y},\ \ \hat{\bf b}=(B^TB)^{-1}B^T({\bf Y}-{\bf Z}\hat{\beta}),\]
	where
	${\bf Y}=(Y_1,\dots,Y_n)^T$, ${\bf Z}=({\bf Z}_1,\dots,{\bf Z}_n)^T$, $B=[<X_i,B_j>]_{1\leq i\leq n}^{1\leq j\leq k_n}$, $A=B(B^TB)^{-1}B^T$. The estimators of $\alpha(t)$ and $\sigma^2$ are
	\[\hat{\alpha}(t)=\sum_{s=1}^{k_n}\hat{b}_sB_s,\ \ \hat{\sigma}^2=\frac{1}{n}\sum(Y_i-<X_i,\hat{\alpha}>-{\bf Z}_i^T\hat{\beta}).\]

	For two positive sequences $\{a_n\}_{n=1}^{+\infty}$ and $\{b_n\}_{n=1}^{+\infty}$, $a_n\asymp b_n$ means $a_n/b_n$ is uniformly bounded away from infinity and zero. To get the asymptotic normality, the following assumptions are required in Zhou et al.(2016). 
	
	(A1) There are positive constants $M$ and $\frac{1}{4(k+1)}<r<\frac{1}{2}$ such that
	\[h=\max_{0\leq j\leq N_n}(t_{j+1}-t_j)\asymp n^{-r},\ k_n\asymp n^{-r},\ h/\min_{0\leq j\leq N_n}(t_{j+1}-t_j)\leq M.\]
	
	(A2) $E||X||^4<+\infty$ and the eigenvalues of the covariance operator $\Gamma$ of $X$ are strictly positive.
	
	(A3) $E|Z_{11}|^4+\dots+E|Z_{1p}|^4+E|\epsilon_1|^4<+\infty$.
	
	(A4) $E(Z_{1j}|X_1)=<X_1,g_j>$ for some function $g_j\in L^2([0,1])$, for $j=1,\dots, p$. Besides, we assume $g_j$, $\alpha\in C^{k+1}([0,1])$.
	
	(A5) Let $\eta_1=(\eta_{11},\dots,\eta_{1p})^T$ with $\eta_{1j}=Z_{1j}-E(Z_{1j}|X_1)$. Assume $\Sigma=Var(\eta_1)$ is positive definite. 
	
	Under conditions (A1)-(A5), Zhou et al. (2016) established the asymptotic normality for the regression parameters and global convergence rate for the slope function as follows.
	
	\begin{lemma} Under condition (A1)-(A5), we have
		\[\sqrt{n}(\hat{\beta}-\beta_0)\Rightarrow N(0,\sigma^2\Sigma^{-1}),\]
		\[||\hat{\alpha}-\alpha||^2=O_p(\frac{k_n}{n}+k_n^{-2(k+1)}),\]
		\[\frac{n}{\hat{\sigma}^2}(\hat{\beta}-\beta_0)^T\hat{\Sigma}^{-1}(\hat{\beta}-\beta_0)\Rightarrow \chi^2_p,\]
		where $\Sigma=Var({\bf Z}_1-E({\bf Z}_1|X_1))$, $"\Rightarrow"$ represents convergence in law and
		\[\hat{\Sigma}=\frac{1}{n}{\bf Z}^T(I-A){\bf Z}.\]
	\end{lemma}

	Then the asymptotic $(1-\gamma)\%$ confidence region is
	\[\bigg\{\beta:\frac{n}{\hat{\sigma}^2}(\hat{\beta}-\beta)^T\hat{\Sigma}^{-1}(\hat{\beta}-\beta)\leq \chi^2_p(1-\gamma)\bigg\}.\]

	\section{The empirical likelihood method}

	Motivated by the estimation equation, we define
	
	\[{\bf W}_i(\beta)={\bf Z}_i\bigg (Y_i-{\bf Z}_i^T\beta-B_i^{*T}(B^TB)^{-1}B^T(Y-{\bf Z}\beta)\bigg ),\]
	where $B_i^{*T}$ is the $i$-th row of the matrix $B$. The empirical likelihood at $\beta$ is given by
	\[\mathcal{R}_n(\beta)=\sup\bigg\{\prod_{i=1}^nn\pi_i|\sum_{i=1}^n\pi_i{\bf W}_i=0,\ \sum_{i=1}^n\pi_i=1,\ \pi_i\geq0\bigg\}.\]
	
	By Lagrange method as in Owen(2001), the solution is
	\[\pi_i=\frac{1}{n}\frac{1}{1+\lambda^T{\bf W}_i},\]
	where $\lambda$ satisfies
	\begin{equation}\label{sol}
	\frac{1}{n}\sum_{i=1}^n\frac{{\bf W}_i}{1+\lambda^T{\bf W}_i}=0.
	\end{equation}

	\begin{theorem}
	Suppose (A1)-(A5) hold. Then at the true $\beta_0$, we have
		\[-2\log\mathcal{R}_n(\beta_0)\Rightarrow {\bf U}^T\Sigma_0{\bf U},\]
	where $"\Rightarrow"$ means "convergence in distribution", ${\bf U}\sim\mathcal{N}(0,I_p)$, $\Sigma_0=\Sigma^{\frac{1}{2}}\Sigma_1^{-1}\Sigma^{\frac{1}{2}}$, $\Sigma=Var({\bf Z}_1-E({\bf Z}_1|X_1))$ and $\Sigma_1=E({\bf Z}_1{\bf Z}_1^T)$.
	\end{theorem}
	The limiting distribution of the empirical likelihood ratio is not the usual chi-square distribution. Actually, it is a weighted sum of independent chi-square distributions with degree of freedom 1. However, the confidence
	regions based on the this empirical likelihood ratio still have the advantages of
   having natural shape
	and respecting the range of $\beta$. In application, $\Sigma$ and $\Sigma_1$ are unknown and we propose the following consistent estimators
	\begin{equation}\label{vest}
	\hat{\Sigma}=\frac{1}{n}\sum_{i=1}^n{\bf Z}_i{\bf Z}_i^T,\ \ \hat{\Sigma}_1=\frac{1}{n}{\bf Z}^T(I-A){\bf Z}.
	\end{equation}
	To construct confidence regions, we use the Monte Carlo method to simulate the limiting distribution and compute the desired quantile.

	\section{Simulation study}
	In this section, we present simulation results to compare the finite sample behaviors of empirical likelihood(EL) method and the asymptotic normality(NA) method in terms of covarage probability of the confidence regions. 
	As in Lian (2011), we generate $X_i$ by the following method,
	\[X_i=\sum_{j=1}^{50}\xi_{ij}j^{-1}\phi_j(t),\]
	where $\phi_1(t)=1$, $\phi_j(t)=\sqrt{2}\cos((j-1)\pi t)$ for $j\geq2$ and $\xi_{ij}$ are iid with uniform distribution $U[-\sqrt{3},\sqrt{3}]$. We simulate data based on the following three models, which are similar to  the models in Zhou et al.(2016).
	
	{\bf Model 1}: $Y_i=Z_{i1}+Z_{i2}+\int_{0}^{1}X_i(t)\alpha(t)dt+\epsilon_i$, $\alpha(t)=\frac{\sqrt{2}}{2}+\sum_{j=2}^{50}4j^{-2}\phi_j(t)$. $(Z_{i1}, Z_{i2})$ are from bivariate normal with zero mean vector, $Var(Z_{i1})=0.9$, $Var(Z_{i2})=0.5$ and $Cov(Z_{i1}, Z_{i2})=0.2$. The error follows $\mathcal{N}(0,0.36)$ and skewed normal distribution with mean 0, standard deviation 1 and skewness parameter 5.

	{\bf Model 2}: $Y_i=5Z_{i1}-1.7Z_{i2}+\int_{0}^{1}X_i(t)\alpha(t)dt+\epsilon_i$. We generate $(Z_{i1}, Z_{i2})$ from standard bivariate normal distribution and use the following functional coefficient,
	\[\alpha(t)=2\sin(0.5\pi t)+4\sin(1.5\pi t)+5\sin(2.5\pi t).\]
	The error follows $\mathcal{N}(0,1)$ and skewed normal distribution with mean 0, standard deviation 1 and skewness parameter 5.

	{\bf Model 3}: $Y_i=2Z_{i1}-Z_{i2}+\int_0^1X_i(t)\alpha(t)+\epsilon_i$, where
	
	\[Z_{i1}=\int_0^1X_i(t)\alpha_1(t)dt+\epsilon_{i1},\ \ Z_{i2}=\int_0^1X_i(t)\alpha_2(t)dt+
	\epsilon_{i2},\]
	$\alpha_1(t)=\sum_{j=1}^{50}b_{1j}\phi_j(t)$, $\alpha_2(t)=\sum_{j=1}^{50}b_{2j}\phi_j(t)$, $b_{11}=1$, $b_{21}=-0.5$, $b_{1j}=2j^{-2}$, $b_{2j}=3j^{-2}$ for $j\geq2$. The random error $\epsilon_{i1}\sim \mathcal{N}(0, 0.25)$ and $\epsilon_{i2}\sim \mathcal{N}(0,0.64)$. The random model error $\epsilon_i\sim \mathcal{N}(0, 0.25)$ and skewed normal distribution with mean 0, standard deviation 0.5 and skewness parameter 5. 
	 
	We use B-spline basis with equally spaced knots and degree 2. The number of knots are selected by the "leave-one-out" cross-validation(Rice and Silverman 1991). We compute the confidence regions for $\beta$ based on the EL method and the NA method. To simulate the limiting distribution in Theorem 1, we firstly estimate the unknown variance and second moment by (\ref{vest}) and then use Monte Carlo method to simulate the $(1-\gamma)$ quantiles. The sample sizes are 30, 50, 80 and 150, representing small, moderate and large sample sizes. The coverage probability is computed by 1000 simulation runs, with nominal confidence level 0.90 and 0.95 respectively. 
	
	The simulation results under Model 1 are summarized in Table 1, with results under Model 2 and Model 3 in Table 2 and Table 3 respectively. There is a similar pattern in all three tables. As the sample size increases, the coverage probabilities increase towards the nominal level, regardless of the error type. However, the EL method outperforms the NA method, since the coverage probabilities based on EL method is larger and closer to the nominal level than that of the NA method. Hence, the EL method yields more accurate confidence regions than the NA method.

		\begin{table}
			\caption{Coverage Probabilities Under Model 1}
			\begin{tabular}{ |p{2.5cm}|p{2cm}|p{2cm}|p{2cm}|p{2cm}|p{2cm}| }
				
				\hline
				\multicolumn{2}{|c|}{ }&\multicolumn{2}{|c|}{Normal Error}&\multicolumn{2}{|c|}{Skewed Normal Error} \\
				\hline
				Sample size $n$    & $1-\gamma$ &EL&NA  &EL&NA\\ 
				\hline
				30   & 0.90  &0.793  &0.760   &0.779  &0.763 \\
				&   0.95     &0.867  &0.841   &0.854  &0.825\\
				\hline
				50 &0.90     &0.840  &0.830   &0.831  &0.822\\
				&0.95        &0.909  &0.897   &0.895  &0.890\\
				\hline
				80&   0.90   &0.843  &0.843   &0.862  &0.853\\
				& 0.95       &0.914  &0.906   &0.922  &0.915\\
				\hline
				150& 0.90    &0.895  &0.889   &0.902  &0.909\\
				&0.95        &0.945  &0.934   &0.952  &0.962\\
				\hline
			\end{tabular}
			
		\end{table}

	\begin{table}
			\caption{Coverage Probabilities Under Model 2}
	\begin{tabular}{ |p{2.5cm}|p{2cm}|p{2cm}|p{2cm}|p{2cm}|p{2cm}| }
	
		\hline
		\multicolumn{2}{|c|}{ }&\multicolumn{2}{|c|}{Normal Error}&\multicolumn{2}{|c|}{Skewed Normal Error} \\
		\hline
		Sample size $n$    & $1-\gamma$ &EL&NA  &EL&NA\\ 
		\hline
		30   & 0.90  &0.776&  0.752 &0.770 &0.746\\
		     & 0.95  & 0.850   &0.819&0.846 &0.820\\
		 \hline
		50   & 0.90 & 0.831&  0.818&0.821 &0.802\\
		     & 0.95 &0.907& 0.897&0.888 &0.876\\
		     \hline
		80   & 0.90  &0.875&0.875&0.860 &0.858\\
		     & 0.95  &0.934   &0.929&0.914 &0.909\\
		\hline
		150  & 0.90  & 0.889&0.883&0.874 &0.867\\
		     &0.95 &0.939 &0.937 &0.925 &0.917\\
		\hline
	\end{tabular}

	\end{table}

		\begin{table}
			\caption{Coverage Probabilities Under Model 3}
			\begin{tabular}{ |p{2.5cm}|p{2cm}|p{2cm}|p{2cm}|p{2cm}|p{2cm}| }
				
				\hline
				\multicolumn{2}{|c|}{ }&\multicolumn{2}{|c|}{Normal Error}&\multicolumn{2}{|c|}{Skewed Normal Error} \\
				\hline
				Sample size $n$    & $1-\gamma$ &EL&NA  &EL&NA\\ 
				\hline
				30   &  0.90    &0.770 &0.744   &0.786 &0.760\\
			    	 &  0.95    &0.839 &0.826   &0.832 &0.816\\
				\hline
				50 &0.90 &0.841 &0.829   &0.837 &0.834\\
				   &0.95 &0.906 &0.893       &0.879 &0.877\\
				\hline
				80  & 0.90  &0.873 &0.862   &0.859 &0.856\\
			    	& 0.95  &0.926 &0.921  &0.918 &0.915\\
				\hline
				150 & 0.90  &0.889 &0.888   &0.879 &0.874\\
			    	& 0.95  &0.941 &0.927  &0.933 &0.919 \\
				\hline
			\end{tabular}
			
		\end{table}

	\section{Proof}
	In this section, we present the proof of Theorem 1.
	
	\begin{lemma} Under conditions (A1)-(A5), we have
		\[\frac{1}{\sqrt{n}}\sum_{i=1}^n{\bf W}_i(\beta_0)\Rightarrow \mathcal{N}(0,\sigma^2\Sigma^{-1}).\]
	\end{lemma}
	
	{\bf Proof:} Note that
	\begin{eqnarray*}
	\frac{1}{\sqrt{n}}\sum_{i=1}^n{\bf W}_i&=&\frac{1}{\sqrt{n}}\sum_{i=1}^n{\bf Z}_i\bigg (Y_i-{\bf Z}_i^T\beta-B_i^{*T}(B^TB)^{-1}B^T(Y-{\bf Z}\beta)\bigg )\\
	&=&\frac{1}{\sqrt{n}}{\bf Z}\bigg (Y-{\bf Z}\beta-B^{T}(B^TB)^{-1}B^T(Y-{\bf Z}\beta)\bigg )\\
	&=&	\frac{1}{\sqrt{n}}{\bf Z}\bigg ((I-A)Y-(I-A){\bf Z}\beta\bigg )\\
	&=&\frac{1}{\sqrt{n}}{\bf Z}(I-A){\bf Z}^T\bigg(({\bf Z}(I-A){\bf Z}^T)^{-1}{\bf Z}(I-A)Y-\beta\bigg)\\
	&=&\frac{1}{n}{\bf Z}(I-A){\bf Z}^T\sqrt{n}(\hat{\beta}-\beta).
	\end{eqnarray*}

	By the Lemma 2 and Theorem 1 in Zhou et al. (2016), the last term converges to $\mathcal{N}(0,\sigma^2\Sigma^{-1})$. 

	\qed

	\begin{lemma}Under conditions (A1)-(A5), we have
	\[	\frac{1}{n}\sum_{i=1}^n{\bf W}_i(\beta_0){\bf W}_i(\beta_0)^T\xrightarrow{P} \sigma^2\Sigma_1.\]
	\end{lemma}

	{\bf Proof:} Firstly, we rewrite the $W_i$ below

	\begin{eqnarray*}
	{\bf W}_i(\beta)&=&{\bf Z}_i\bigg (Y_i-{\bf Z}_i^T\beta-B_i^{*T}(B^TB)^{-1}B^T(Y-{\bf Z}\beta)\bigg )\\
	&=&{\bf Z}_i\bigg (Y_i-{\bf Z}_i^T\beta-B_i^{*T}(B^TB)^{-1}B^T(Y-{\bf Z}\hat{\beta}+{\bf Z}(\hat{\beta}-\beta))\bigg )\\
	&=&{\bf Z}_i\bigg (Y_i-{\bf Z}_i^T\beta-B_i^{*T}\hat{b}-B_i^{*T}(B^TB)^{-1}B^T{\bf Z}(\hat{\beta}-\beta)\bigg )\\
	&=&{\bf Z}_i\bigg (Y_i-{\bf Z}_i^T\beta-<X_i,\hat{\alpha}>-B_i^{*T}(B^TB)^{-1}B^T{\bf Z}(\hat{\beta}-\beta)\bigg )\\
	&=&{\bf Z}_i\bigg (Y_i-{\bf Z}_i^T\beta-<X_i,\hat{\alpha}>-B_i^{*T}(B^TB)^{-1}B^T{\bf Z}(\hat{\beta}-\beta)\bigg )\\
	&=&{\bf Z}_i\bigg (Y_i-{\bf Z}_i^T\beta-<X_i,\alpha>-<X_i,\hat{\alpha}-\alpha>-B_i^{*T}(B^TB)^{-1}B^T{\bf Z}(\hat{\beta}-\beta)\bigg )\\
	&=&{\bf Z}_i\bigg (\epsilon_i-<X_i,\hat{\alpha}-\alpha>-B_i^{*T}(B^TB)^{-1}B^T{\bf Z}(\hat{\beta}-\beta)\bigg ).
	\end{eqnarray*}

	Then 
	
	\begin{eqnarray*}
		\frac{1}{n}\sum_{i=1}^n{\bf W}_i{\bf W}_i^T&=&\frac{1}{n}\sum_{i=1}^n{\bf Z}_i{\bf Z}_i^T\bigg\{\epsilon_i^2+<X_i,\hat{\alpha}-\alpha>^2-2<X_i,\hat{\alpha}-\alpha>\epsilon_i\\
		& &+B_i^{*T}(B^TB)^{-1}B^TZ(\hat{\beta}-\beta)(\hat{\beta}-\beta)^TZ^T(B^TB)^{-1}B_i^{*}\\
		& &-2 B_i^{*T}(B^TB)^{-1}B^TZ(\hat{\beta}-\beta)\epsilon_i\\
		& &+2 B_i^{*T}(B^TB)^{-1}B^TZ(\hat{\beta}-\beta)<X_i,\hat{\alpha}-\alpha>\bigg\}\\
		&=&R_{n1}+R_{n2}+R_{n3}+R_{n4}+R_{n5}+R_{n6}.
	\end{eqnarray*}

	By the law of large number and the fact $\epsilon_i$ and ${\bf Z}_i$ are independent, we have
	\[R_{n1}=\frac{1}{n}\sum_{i=1}^n{\bf Z}_i{\bf Z}_i^T\epsilon_i^2\xrightarrow{P}E(\epsilon_1^2{\bf Z}_1{\bf Z}_1^T)= \sigma^2\Sigma_1.\]

	Next, we show the rest terms converge to zero in probability.
	
	\[E<X_i,\hat{\alpha}-\alpha>^2\leq E||\hat{\alpha}-\alpha||^2||X_i||^2\leq\sqrt{E||\hat{\alpha}-\alpha||^4E||X_i||^4}\]
	By Lemma 1 and (A2), we have
	\[<X_i,\hat{\alpha}-\alpha>^2=o_p(1),\]
	hence $R_{n2}=o_p(1)$ and $R_{n3}=o_p(1)$.

	Notice that by Theorem 1 and Lemma 1 in Zhang et al. (2016),
	\begin{eqnarray*}
	B_i^{*T}(B^TB)^{-1}B^T{\bf Z}(\hat{\beta}-\beta)&=&\frac{1}{n\sqrt{n}}B_i^{*T}\bigg(\frac{B^TB}{n}\bigg)^{-1}B^T{\bf Z}\sqrt{n}(\hat{\beta}-\beta)\\
	&\asymp&\frac{1}{n\sqrt{n}}B_i^{*T}B^T{\bf Z}.
	\end{eqnarray*}

	By (A2) and (A4), one has
	\begin{eqnarray*}
	E||B_i^{*T}B^T{\bf Z}||_2&=&E\bigg|\bigg|\sum_{l=1}^n\sum_{j=1}^{k_n}<B_j, X_l><B_j,X_i>{\bf Z}_l^T\bigg|\bigg|_2\\
	&\leq&\sum_{l=1}^n\sum_{j=1}^{k_n}E\bigg|\bigg|<B_j, X_l><B_j,X_i>{\bf Z}_l^T\bigg|\bigg|_2\\
	&\leq&\sum_{l=1}^n\sum_{j=1}^{k_n}E||B_j||^2||X_l||\ ||X_i||\ ||{\bf Z}_l^T||_2\\
	&\lesssim&nk_n,
	\end{eqnarray*}
	 where $||.||_2$ is the Euclidean norm.

	Hence,
	\[B_i^{*T}(B^TB)^{-1}B^T{\bf Z}(\hat{\beta}-\beta)=O_p(\frac{k_n}{\sqrt{n}})=o_p(1).\]
	Then $R_{n4}=R_{n5}=R_{n6}=o_p(1)$.

	\qed

	\begin{lemma} Under the assumption (A1)-(A5), one has
		\[\max_{1\leq i\leq n}||{\bf W}_i||_2=o_p(\sqrt{n}),\ \ ||\lambda||_2=O_p(n^{-\frac{1}{2}}).\]
	\end{lemma}

	{\bf Proof:} By the proof of Lemma 3, we have
	\[\epsilon_i-<X_i,\hat{\alpha}-\alpha>-B_i^{*T}(B^TB)^{-1}B^T{\bf Z}(\hat{\beta}-\beta)=O_p(1).\]

	And since $0<Var({\bf Z}_1)<+\infty$, then $\max_{1\leq i\leq n}||{\bf Z}_i||_2=o_p(\sqrt{n})$ by Owen (2001), which completes the proof.
	
	\qed

	{\bf Proof of Theorem 1}: By (\ref{sol}), Lemma 2 and Lemma 4, we have 
	
	\[\lambda(\beta_0)=\bigg (\frac{1}{n}\sum_{i=1}^n{\bf W}_i{\bf W}_i^T\bigg )^{-1}\frac{1}{n}\sum_{i=1}^n{\bf W}_i+O_p(\frac{1}{\sqrt{n}}).\]

	Then by Lemma 3, it yields the following Taylor expansion 
	\begin{eqnarray*}
	-2\log\mathcal{R}_n(\beta_0)&=&\sum_{i=1}^n\lambda(\beta_0){\bf W}_i(\beta_0)+o_p(1)\\
		&=&\bigg (\frac{1}{\sqrt{n}}\sum_{i=1}^n{\bf W}_i\bigg )^T\bigg (\frac{1}{n}\sum_{i=1}^n{\bf W}_i{\bf W}_i^T\bigg )^{-1}\bigg (\frac{1}{\sqrt{n}}\sum_{i=1}^n{\bf W}_i\bigg )+o_p(1)\\
		&\Rightarrow& {\bf G}^T\Sigma_1^{-1}{\bf G},
	\end{eqnarray*}
	where ${\bf G}\sim \mathcal{N}(0,\Sigma^{-1})$. The proof of Theorem 1 is complete.

\end{document}